\documentstyle{article}
\def\ii{\'{\i}}
\begin{document}
\title{Pseudo SU(3) Model and Abnormal Parity States\thanks{Work supported
in part by Conacyt under project 1570-E9208}}
\author{Jorge G. Hirsch \\
  Departamento de F\ii sica\\
 Centro de Investigaci\'on y Estudios Avanzados del IPN\\
 Apdo. Postal 14-740, M\'exico 07000 D.F.
\and
Octavio Casta\~nos and Peter O. Hess \\
Instituto de Ciencias Nucleares\\
Universidad Nacional Aut\'onoma de M\'exico\\
Apdo. Postal 70-543, M\'exico 04510 D.F.}
\date{}
\maketitle
\begin{abstract}
The most important features of the pseudo SU(3) model are reviewed. This
description of heavy deformed nuclei is based on
 pseudo spin symmetry, and is able to correctly predict the collective
spectra and their transition amplitudes, using appropriate effective charges.
It also gives a very good description of the exotic double beta decay, a hard
test of any nuclear model. The simplest pseudo SU(3)
assumptions describe nucleons in the abnormal parity  states as having
seniority zero. This
idea is very useful as a mathematical simplification, and is also able to
give a nice description of the backbending phenomena, but has been challenged
in a recent study about effective charges and $BE2(0^+_1 \rightarrow 2^+_1)$
systematics. We carefully checked this statement and
conclude that the seniority zero approach is a formal way of assign
abnormal parity nucleons a passive role. The contribution of the abnormal
parity nucleons to the nuclear quadrupole moments is finite but only a fraction
of their asymptotic Nilsson values.

\vskip .3cm
\centerline{\bf Resumen}
\vskip .2cm
Se revisan los aspectos m\'as importantes del modelo pseudo SU(3). Esta
descripci\'on de los n\'ucleos pesados deformados est\'a basada en la
simetr\'{\i}a de pseudo-esp\'{\i}n y es capaz de predecir correctamente los
espectros colectivos y sus amplitudes de transici\'on usando cargas
efectivas apropiadas. Tambi\'en proveee una muy buena descripci\'on del
ex\'otico decaimiento beta doble, una dif\'{\i}cil prueba para cualquier
modelo nuclear.
El modelo pseudo SU(3) m\'as sencillo describe a los nucleones en los
estados de paridad anormal teniendo antig\"uedad cero. Esta es una
simplificaci\'on matem\'atica muy \'util y puede dar una aceptable
descripci\'on del ``backbending'', pero recientes estudios sobra cargas
efectivas e intensidades $BE2(0^+_1 \rightarrow 2^+_1)$ la han
cuestionado. Nosotros revisamos este postulado y concluimos que la
aproximaci\'on de antig\"uedad cero es una manera formal de asignar a los
nucleones con paridad anormal un rol pasivo. La contribuci\'on de estos
nucleones a los momentos cuadrupolares nucleares es finita, pero solo una
fracci\'on de sus valores asint\'oticos.
 \end{abstract}


\begin{section}{Introduction}

The pseudo SU(3) model has been developed during the last two decades
\cite{Rat73,Rat79,Dra81,Dra84}. Its
main aim is to extend the successful application of the SU(3) symmetry from
light to heavy deformed nuclei. This possibility is based in the
existence of the pseudo spin symmetry
\cite{Ari69}. Based on this concept, it was possible
to construct a transformation from the real deformed single particle states,
where the spin-orbit interaction completely destroys the three dimensional
harmonic oscillator structure of the single particle spectra present in
light nuclei, to the pseudo SU(3) states, where the ``pseudo'' spin
orbit interaction is negligible, and the underlying SU(3) symmetry is
recovered for normal parity states.

Having this new SU(3) symmetry, it has been found that many important
features of heavy deformed nuclei can be reproduced. For example, the
spectra and
E2, M1, M3 transition amplitudes for some rare earth and actinides
were correctly predicted
using appropriate effective charges\cite{Dra84,Cas87}.
Also the exotic
 double beta decay half lives of some nuclei were predicted with notably
accuracy\cite{Hir92}.

These large capabilities of the pseudo SU(3) model
lead us to the conclusion that
the pseudo SU(3) wave functions have a significant overlap with the exact
wave functions of heavy deformed nuclei,
or at least are the most important part in the phenomena described with
the model. Core polarisation, as in  shell model
calculations, is simulated using effective charges.

Single particle wave functions, both in deformed Nilsson or Woods Saxon
potentials, exhibit the intruder states of abnormal parity as nearly unmixed
for normal deformations ($\beta \leq 0.3$
\cite{Rin80,Ber89}. In this limited single-j
subspace pairing strongly competes with the quadrupole-quadrupole interaction.
These unique parity nucleons were assumed to have seniority zero,
providing a framework in which the whole wave function
is of the SU(3) type, an assumption that greatly  simplifies
calculations.

A comparison of effective charges needed in the pseudo SU(3) and the
Single Shell Asymptotic Nilsson Model (SSANM) exhibits the limitations
of the seniority zero approach\cite{Bha92}.
It shows that effective charges must be
renormalized by nearly 25\% to compensate for the lack of quadrupole
moment associated with the nucleons in the abnormal parity states.

In this paper we examine again some relevant characteristics and predictions
of the pseudo SU(3) model. In particular, a possible description of the
backbending phenomena based on the seniority zero approach is reviewed
\cite{Rat79}.
We present a discussion of the quadrupole
moments and $BE2\uparrow \equiv BE2(0^+_1 \rightarrow 2^+_1)$ intensities in
the pseudo SU(3) and SSANM models. We confirm the results found in
\cite{Bha92},
that the quadrupole moments in the normal parity sector are the same in both
models. We also discuss several predictions for the quadrupole moments of
the abnormal parity nucleons.

\begin{section}{The pseudo SU(3) model}

A tractable shell model theory of heavy deformed nuclei requires a
severe truncation of the spherical model basis. Of course the goal of
truncation, to reproduce the essential physics found in
low-lying states of a larger space in a smaller one, can only be
achieved if the basis selection is made relative to those parts of the
interaction that dominate the low-energy structure\cite{Dra84}.

The organization of the basis states relies on uncovering an SU(3)
symmetry in the structure of higher major shells, where
 SU(3) is the symmetry group of the three-dimensional harmonic
oscillator (HO). Since for lower {\em ds}-shell nuclei
the nuclear shell structure is not much different
from that of the HO, SU(3) was proposed
by Elliott\cite{Ell58} as a reasonable {\em ds}-shell symmetry, a one that
could be used to truncate the full space($10^{3-5}$ states) down to
tractable
size ($10^{1-2}$ states). Since $Q\cdot Q$ is dominant and $Q\cdot Q =
4C_2
-3L\cdot L$, where $C_2$ is the second order Casimir invariant of SU(3),
irreducible representations (irreps) of SU(3) which have the largest
values of $C_2$ should dominate the structure of low-lying states. In
nuclear physics SU(3) irreps are labeled by $(\lambda , \mu )$\cite{Dra84} and
\begin{equation}
<C_2>= (\lambda + \mu +3)(\lambda + \mu) - \lambda \mu \hspace{1cm} .
\label{c2}
\end{equation}
 Basis states
belonging to this "leading" irrep of SU(3) are those which have the
largest intrinsic quadrupole deformation, $<Q_0 >\approx <Q\cdot
Q>^{1/2}$. A severe truncation scheme would restrict basis states to the
leading irrep. Full space {\em ds}-shell calculations have confirmed
that the leading irreps do indeed comprise $60-80\%$ of the yrast
(lowest state of a given spin) eigenstates.

For the higher major shells the spin orbit and centrifugal streching
perturbations completely destroy the HO shell structure. In all cases,
however, there are new major shells. These new major shells are
comprised of all the $j$ subshells of the corresponding HO major shell
except for the largest $j$ subshell which is pushed into the next lower
major shell. The remaining $j$ subshells are grouped together and called
the normal (N) parity orbitals of the new shell. In addition to these,
there is the highest $j$ subshell from the next higher HO shell. This
``intruder" level has the parity opposite to the other levels in the
major shell and is called the abnormal (A) (or unique) parity level. As
an example, consider the $\eta =5$ major shell of the HO. It has the
degenerate subshells $h_{11/2}, h_{9/2}, f_{7/2}, f_{5/2}, p_{3/2},
p_{1/2}$. The corresponding nuclear major shell, however, has the
subshells $ h_{9/2}, f_{7/2}, f_{5/2}, p_{3/2},p_{1/2}$ and the abnormal parity
$i_{13/2}$ level with the splittings shown in Fig.1 (deformation
$\epsilon = 0$) for protons. This breakdown of the HO structure means
that the ``real" SU(3) symmetry is not expected to be good even though
$Q\cdot Q$ remains a dominant interaction.

Now consider the normal parity set of orbits; namely, the
$ h_{9/2}, f_{7/2}, f_{5/2}, p_{3/2}$ and $p_{1/2}$ that come from the
$\eta =5$ HO shell. Note that the $j$ values (ignore $l$ for the moment)
are
exactly the $j$ values which appear in the $\eta =4$ HO shell. This
suggest
that one can map the {\bf l} and {\bf s} (preserve {\bf j}) for these
levels onto
\~{\bf l} and \~{\bf s} , i.e, {\bf j} = {\bf l} + {\bf s} $\rightarrow$
\~{\bf l} + \~{\bf s} = {\bf j} such that
$ h_{9/2}, f_{7/2}, f_{5/2}, p_{3/2},p_{1/2} \rightarrow \tilde
g_{9/2},\tilde g_{7/2}, \tilde d_{5/2}, \tilde d_{3/2}, \tilde s_{1/2}$.
This mapping scheme grew out of the studies of pseudo spin-orbit
doublets\cite{Ari69}. It has been used by Raju, Draayer and Hecht
\cite{Rat73} in the discussion of the ground state magnetic moments of
odd-A deformed nuclei and by Strottman \cite {Str72} in a study of
Ni-Cu-Zn isotopes. Hecht \cite{Hec75} has used the scheme to investigate
decoupled negative parity spectra of the Au nuclei while Braunschweig
and Hecht \cite{Bra78} used it in an analysis of the core deformation
for nuclei in the proton rich Xe-Nd region. The scheme has also been
used in attempts at providing a microscopic shell-model interpretation
of high-spin phenomena in Ge and Ba nuclei\cite{Dra81}.
Using an effective interaction comprised of operators
which form an integrity basis for the $SU(3) \rightarrow R(3)$ algebra
was shown to be sufficient to reproduce almost exactly, within a single
leading irreducible representation of SU(3), the ground and gamma band
rotational structure of eight rare earth and four actinide nuclei,
reproducing accurately the interband and intraband E2 strength
\cite{Dra84}. A rigorous test of the theory was provided by the
prediction, in essentially the same nuclei, of a number of $1^+$
states with strong M1 transitions to ground state, as well as E2 and M3
transition strengths, using the real M1, E2 and M3
operators\cite{Cas87}. It was also possible to describe the two neutrino mode
of the double beta decay in $^{150}Nd$ and $^{238}U$ in good agreement with
the experimental values\cite{Hir92}. This latter prediction is far from
trivial,
given this exotic decay has challenged nuclear structure descriptions for more
than a decade.

Recently an analytic expression for the
transformation
that take us from the normal parity orbitals to the pseudo-space was
introduced\cite{Cas92}. By applying this transformation to the spherical
Nilsson
Hamiltonian it can be shown explicitly that the strength of the pseudo
spin-orbit interaction is almost zero and the orbitals $j =\tilde l \pm
1/2$ are nearly degenerate doublets \cite{Dra84,Dra82,Naq90}.
The relabeled levels form a major shell for the pseudo oscillator
potential. The symmetry of this oscillator is, of course,
pseudo SU(3)
(the ``pseudo'' denotes the pseudo-shell realization but the abstract algebra
is just SU(3).) This ``real" to ``pseudo" shell mapping is a (restricted)
unitary transformation and does not change anything as far as exact full
space results are concerned.
A tensor decomposition of the real $Q\cdot Q$ interaction into its
pseudo SU(3) components shows that it is predominantly $\tilde Q \cdot
\tilde Q$, i.e. the quadrupole-quadrupole interaction in the pseudo
shell. Also, although it is the spin-orbit interaction that destroys the
HO shell structure, note the behavior of the Nilsson levels labeled by
their $\tilde\Omega [\tilde\eta \tilde n_z \tilde\Lambda ]$ quantum
numbers in Fig. 1. For deformations appropriate to well-deformed nuclei,
$\epsilon \approx 0.3$ ( with $\epsilon = 0.95 \beta$),
one sees that the levels are grouped into pseudo
spin-orbit doublets ($\tilde\Omega = \tilde\Lambda \pm 1/2$). This
implies that the pseudo spin-orbit splitting is small, supporting the
assumptions of the model, and the formal results for the transformation
\cite{Cas92}. Similarly, the equivalent of the centrifugal streching
interaction which spreads the \~{$l$} values is less strong here than
in the HO picture. Also note that adding nucleons to the closed shell
tends to create an intrinsic state with the largest possible total
$\tilde n_z$ (for fixed deformation in prolate deformed nuclei, $\tilde
n_z$ decreases
as energy increases) and hence the largest quadrupole momentum,

\begin{equation}
<\tilde Q_0> \approx  2 \tilde n_z - \tilde n_x - \tilde n_y = 3 \tilde
n_z
- \tilde\eta \hspace{1cm} .
\end{equation}

An additional complication for heavy nuclei is that the valence protons
($\pi$) and neutrons ($\nu$) are filling different majors shells. Thus
for a given nucleus there are two open shells, one for protons and one
for neutrons, each comprised of a set of normal parity levels and the
associated abnormal parity level. Since the nuclear interaction is
assumed to conserve parity as well as the numbers of protons and
neutrons, one might be led to consider a basis built by weak coupling
configurations of the four separate spaces. However, in a previous study
\cite{Dra82} it was found that if the residual interaction contains a
significant $Q_{\pi} \cdot Q_{\nu}$ part, then coupling the leading
proton pseudo SU(3) configuration to the leading neutron pseudo SU(3)
configuration leads to an SU(3) scheme whose leading irreps dominate the
low-lying structure. To this strong coupled space we couple weakly the
remaining abnormal parity spaces.

\begin{section}{The Abnormal Parity States}

In the abnormal parity parts of the neutron and proton shells it was assumed
in previous work\cite{Dra84,Cas87} that low seniority configurations were
the most important ones. Thus, only seniority zero configurations were
taken into account.

In order to understand this seniority zero postulate, we will review in some
detail
the microscopical description of the backbending phenomenon. It has been
explained by Stephens and Simon\cite{Rin80,Ste72} as an alignment of two
neutrons in the intruder $i_{13/2}$ shell. If these two neutrons,
instead of rotating around the 3-axis, align along the rotational axis
of the nucleus, this adds and additional $13/2 + 11/2 = 12$ units of
angular momentum. Therefore, the nucleus can decrease its collective
rotation while increasing its total angular momentum through the
addition of single-particle angular momentum.

These are the basic ideas which support a microscopic look at backbending,
in a work done by Ratna Raju, Hecht, Chang and Draayer\cite{Rat79}. The
rotational core is naturally provided by the normal protons and neutrons
strongly coupled in the pseudo SU(3) basis. The abnormal parity
$h_{11/2}$ orbital provides the seniority zero and two states, and
the observed backbending in $^{126}Ba$ is reproduced. In Fig. 2
their results are displayed.
They selected the irrep $(\tilde\lambda ,\tilde\mu ) = (24,0)$ for
the $\tilde f,\tilde p$ pseudo-shell  for both protons and neutrons. Its
spectrum is shown in the right hand side of Fig. 2, and having been obtained
for a Hamiltonian with a surface delta interaction, it resembles very
much a rigid rotor limit (pure $Q\cdot Q$ interaction).
The first column of the far right hand side shows the spectrum obtained for
the particles in the pure subspace of the abnormal $h_{11/2}$ level,
with one seniority zero
state and five seniority two states, coupled to I=2,4,6,8,10, which are
not degenerated because of the presence of a Hamiltonian with a surface
delta interaction in the abnormal sector. The Hilbert space is
constructed as the direct product of the normal and abnormal subspaces,
coupled to good angular momentum, and good isospin. An interaction
Hamiltonian able to mix the normal and abnormal spaces is introduced.
The resulting spectrum of the composite product space is shown in
the middle, and in the left hand side are the experimental results. The
backbending of the yrast band between spins I=10-14 is obvious to the
trained eye. The relative intensities of the three interaction channels
(normal, abnormal and mixed) were selected to reproduce the
experimental information. The mixing channel must be low enough to
allow the backbending effect to occur. If the mixing is too strong, the
rotational and seniority two states would become mixed such that no
sudden change in the ground band, i.e. backbending, can be seen.

This microscopical description of the backbending phenomenon provides
a plausible picture of the role played by the particles in
the abnormal parity states. But it must be stressed that this is not the most
usual description of the backbending
phenomenon. Although the main ideas are the same as
in other models, there exist a crucial difference: the pairs of abnormal
parity nucleons are described as seniority zero ones, i.e. they are coupled
to angular momentum zero {\em in the spherical basis}. The most common
approach is to fill Nilsson levels with pairs of nucleons with {\em angular
momentum projection zero}, and then to project these states in order
to obtain states with good total angular momentum \cite{Rin80,Ste72}. These
pairs
in the deformed basis could have partial occupation, being then described
as quasiparticles. But even in this case the {\em deformed} zero quasiparticle
state does not coincide with the seniority zero one.

\section{Quadrupole Moments}

Now we give a more detailed look to the quadrupole transition strength.
The reduced electric quadrupole transition intensity
$BE2\uparrow$ is given in terms of the electric quadrupole
moment Q\cite{Bha92} by
\begin{equation}
BE2\uparrow = \frac 5 {16\pi} |Q|^2
\end{equation}
with
\begin{equation}
Q = e_\pi Q_\pi + e_\nu Q_\nu
\end{equation}
and
\begin{equation}
\begin{array}{ll}
Q_\alpha &= <2^+M| q^M_\alpha |0^+0>\\
 &= \sqrt{16 \pi / 5} \sum_i <2^+M|r^2_\alpha(i)
Y_{2M}(\hbox{\^r}_\alpha (i))|0^+0>, \hspace{.4cm}\alpha =\pi ,\nu
\end{array}
\end{equation}

\noindent
where $Q_{\alpha}$ represents the quadrupole transition amplitude, and in
this case is essentially the same as the intrinsic quadrupole moment.

In the above expressions the real quadrupole operator appears.
In order to evaluate them
 between pseudo SU(3) states, it must be expanded in the
pseudo SU(3) basis, as explained in \cite{Cas87}. In Table II of this
reference the explicit expansion of the quadrupole operator in its
pseudo SU(3) components is shown, and it becomes evident that Q is
proportional to
\~Q, given the $(\lambda ,\mu) = (1,1), K=1, L=2,S=0$ component has
largely the greatest component. But there exist a numerical factor,
associated with the fact that the pseudo shell has one phonon less than
the real one, and, as can be checked from this table II, it can be
expressed approximately by
\begin{equation}
 Q_\alpha \approx \frac {\eta_\alpha +1} {\eta_\alpha} \tilde Q_\alpha = \frac
{\tilde\eta_\alpha + 2}
{\tilde\eta_\alpha +1} \tilde Q_\alpha \hspace{1cm} .
\end{equation}

It is important to notice that the $|0^+>$ and $|2^+>$ states are
strongly
coupled states in the pseudo SU(3) basis. They are labeled by their
total
$(\lambda , \mu )$, their orbital angular momentum L with projection M,
their spin S=0 (which gives J=L), and an additional label K
distinguishing the different rotational bands.
 The final expression for
$\tilde Q_\alpha$ becomes
\begin{equation}
\tilde Q_\alpha = <(\lambda_\pi ,\mu_\pi ) (\lambda_\nu ,\mu_\nu )
(\lambda
,\mu)1 2 M | q^M_\alpha |(\lambda_\pi ,\mu_\pi ) (\lambda_\nu ,\mu_\nu )
(\lambda ,\mu ) 1 0 0>
\end{equation}
where, for the leading strong coupled irrep
\begin{equation}
\lambda = \lambda_\pi + \lambda_\nu ; \hspace{1cm} \mu = \mu_\pi +
\mu_\nu  \hspace{1cm} .
\end{equation}
The final expressions to be evaluated are\cite{Dra84,Cas87}
\begin{eqnarray}
\tilde Q_\pi = \sqrt{\frac 5 {16\pi}}
\left\{ \begin{array}{cccc}
        (\lambda_\pi ,\mu_\pi) &(1,1) &(\lambda_\pi ,\mu_\pi) &1 \\
        (\lambda_\nu ,\mu_\nu) &(0,0) &(\lambda_\nu ,\mu_\nu) &1 \\
        (\lambda ,\mu ) &(1,1) &(\lambda ,\mu ) &1 \\
        1               &1     &1
        \end{array} \right\} \\
<(\lambda ,\mu ) 10 ,(1,1)12 \|(\lambda , \mu ) 12>_{\rho=1}
\sqrt{4 C_2(\lambda_\pi ,\mu_\pi )} \nonumber
\end{eqnarray}
 and
\begin{eqnarray}
\tilde Q_\nu = \sqrt{\frac 5 {16\pi}}
\left\{ \begin{array}{cccc}
        (\lambda_\pi ,\mu_\pi) &(0,0) &(\lambda_\pi ,\mu_\pi) &1 \\
        (\lambda_\nu ,\mu_\nu) &(1,1) &(\lambda_\nu ,\mu_\nu) &1 \\
        (\lambda ,\mu ) &(1,1) &(\lambda ,\mu ) &1 \\
        1               &1     &1
        \end{array} \right\} \\
<(\lambda ,\mu ) 10 ,(1,1)12 \|(\lambda , \mu ) 12>_{\rho=1}
\sqrt{4 C_2(\lambda_\nu ,\mu_\nu )} \nonumber
\end{eqnarray}
The above expressions seems sligthly involved, but they are
easy
to evaluate given the $\{ ... \}  (9-\lambda \mu$ coefficient) and the
$<...,...\|...>$ (SU(3) reduced Clebsch-Gordan coefficient) are well known
quantities, and there are available computer codes \cite{Dra73}
which make their use nearly
as easy as the more known R(3) Clebsch-Gordan coefficients.  The
explicit expression for $C_2$ is given in eq.(\ref{c2}). In Table I we
present, for twelve heavy deformed nuclei, their number of protons and
neutrons in the normal parity states ($n^N_\pi$ and $n^N_\nu $), their
irreps $(\lambda_\pi ,\mu_\pi)$ and $(\lambda_\nu ,\mu_\nu)$, the
quadrupole transition amplitudes $Q_\pi$ and $Q_\nu$ and, in the last
column, their approximate intrinsic values (in the decoupled basis)
\begin{equation}
Q^{int}_\alpha = \frac {\eta_\alpha + 1} {\eta_\alpha} (2 \lambda_\alpha
+ \mu_\alpha ),\hspace{1cm} \alpha = \pi ,\nu   \label{qint}
\end{equation}

 From Table I it is clear that the intrinsic Q values
represent a very good approximation to the exact ones.

\section{The single shell asymptotic Nilsson \- model (SSANM)}

This model was proposed in Ref. \cite{Ram91} and used extensively in BNR.
Its aim is to give a simple way to calculate intrinsic quadrupole
moments for nuclei in different shells.
Its ansatz is that ``a nucleus is as
deformed as it can be in a single shell".

As an attempt to give some microscopical support to the model, in
Ref.\cite{Bha92,Ram91} it is suggested that the behavior
 of the Nilsson energy levels as a function of deformation
shows that, beyond the deformation of $\beta \approx 0.15$ (or,
equivalently, $\epsilon \approx 0.14$), the slopes of the energy curves
for different Nilsson levels become almost constant for most of the
levels. The slope of the Nilsson energy curve is
proportional to the quadrupole moment of the state. Hence, the constancy
of this slope would imply that the quadrupole moments of the Nilsson
orbits
change very little beyond $\beta \approx 0.15$. Because most deformed
nuclei have larger deformations, it can be assumed that the quadrupole
moments of the various occupied Nilsson orbits would be approximately
constant for different nuclei. The change in the intrinsic quadrupole
moment from one nucleus to another is therefore due to the different number
of valence nucleons occupying the available orbits with the largest
quadrupole moments. The spectrum of the quadrupole moments of the
Nilsson states with large deformation is obtained by considering the
single-particle states in a major shell to be degenerate and then
diagonalizing the $q_0$ operator in that space. Their eigenvalues are
just the mass quadrupole moments of these deformed single-particle
states.

For the normal parity states, we decided, instead of diagonalizing the
above mentioned matrix, to use the pseudo SU(3) formalism described
above. We construct the states in
 the pseudo shell $\tilde\eta = \eta -1$, with asymptotic
quantum numbers $(\tilde\eta,\tilde n_z,\tilde\Lambda)
\Omega = \tilde \Lambda \pm
1/2$ and the parity $\pi = (-1)^\eta$, and their intrinsic quadrupole
moments are given by (\ref{qint}).
The results are exhibited in Table II.
 In the last column the $q_\Omega$ values
of table II of BNR are reproduced, and it is evident that they are
very similar.

We have demonstrated the  BNR assumption
that the SSANM {\em without} the abnormal parity states must give results
analogs to the pseudo SU(3) to a very good approximation. In BNR it is
explained as some kind of ``redefinition" of the quadrupole operator, a
misunderstanding of what is simply a mathematical transformation, but
the conclusion remains valid.

It remains to be discussed the role
of the abnormal parity states. The pseudo SU(3) and the SSANM have
exactly the same assumption (the nucleus is as deformed as it can be)
and the same results in the normal parity sector. Is it correct to give
a similar deformation to the nucleons in the abnormal parity part?

\section{Effective charges}

The distortion of the closed shells by an added particle can be simply
understood as a consequence of the nonspherical field generated by the
extra particle. The order of magnitude of the effect can be estimated by
observing that the eccentricity of the density distribution is of order
$A^{-1}$, and hence, the potential should acquire a similar shape. The
orbit of each proton in the closed shell is thus slightly distorted and
acquires an extra quadrupole moment of the order $A^{-1}~Q_{sp}$ and of
the same sign as the mass quadrupole moment of the polarizing particle.
The total induced quadrupole moment is of the order\cite{Boh69}
\begin{equation}
Q_{pol} \approx \frac Z A  Q_{sp}
\end{equation}
 From this expression it is quite natural to infer
\begin{equation}
e_{pol} = \frac Z A e
\end{equation}

\noindent
with
\begin{equation}
e^{eff}_{\pi} = e + e_{pol},\hspace{2cm} e^{eff}_{\nu} = e_{pol}
\end{equation}
The {\em effective charge} $e^{eff}$ was introduced to reflect
the polarization effect of the core.

This is the approach selected in BNR, with and additional parameter
$\xi > 1$ multipling $e_{pol}$ in the neutron case, to reproduce
the accepted trend that the polarization charge is somewhat larger for a
neutron than for a proton\cite{Boh69}. Their expressions are
\begin{equation}
e^{eff}_{\pi} = e (1 + \frac Z A ). \hspace{2cm}
e^{eff}_{\nu} = \xi e \frac Z A       \label{Ram1}
\end{equation}

This parametrization of the
effective charges is not unique.  For example,
in their classical study of spherical even-even nuclei
using the pairing plus quadrupole hamiltonian, Kisslinger and
Sorensen\cite{Kis63} adopted the parametrization
$e^{eff}_{\pi}=2.0e$ and $e^{eff}_{\nu}=1.0e$, as well as Ring and
Schuck
did in their book (\cite{Rin80}, pages 65 and 389), where a microscopic
description of the effective charges is provided. In some cases,
different sets of $e^{eff}$ are associated with the same nuclei near
closed shells (see Ref.\cite{Rin73} for nuclei near $^{208}Pb$).

In the spirit of trying to compare different models, in order to
obtain a deep understanding of the underlying physics, we will retain
the philosophy postulated in BNR, but propose a different
parametrization, one which will not disqualify the pseudo SU(3) approach
from the beginning, and will allow us to perform a very detailed
analysis of both the SSANM and the pseudo SU(3) model.
This proposal is
\begin{equation}
e_{\pi} = e (1 + \xi Z / A), \hspace{2cm} e_{\nu} = e \xi Z /
A
\label{nos}
\end{equation}
where the parameter $\xi$ can vary from 0 to 4, describing the
situation from the bare proton charge to one strongly correlated with
the core. In this parametrization we loose the slightly increase of
$e^{pol}_{\nu}$ over $e^{pol}_{\pi}$, but we could well say that it is
at least as valid as the other one.

Fig. 3 shows the behavior of
\begin{equation}
C^{model} = Q(exp) / Q(model)
\end{equation}
for the SSANM and the pseudo SU(3). In Fig. 3a the parametrization
(\ref{Ram1}) is used, while in Fig. 3b  the alternative (\ref{nos})
is
employed, in the case of $^{168}Er$. Fig. 3a exhibits the behavior of
Fig. 4 of BNR, and for $\xi = 2.1$ $C^{SSANM} =1$ and $C^{
psSU(3)} \approx 1.5$, and is the basic support for the BNR claim of
a
$50\%$ underestimation of Q in the pseudo SU(3). In Fig. 3a it is
clear, also, that the limit $C^{psSU(3)}=1$ is never achieved,
for the allowed values of $\xi$.
Fig. 3b exhibit the behavior of the same quantities, using the
parametrization (\ref{nos}). In this case, it is clear that both models
are able to reproduce the experimental data, with effective charges
\begin{equation}
\begin{array}{cll}
SSANM: &e^{eff}_{\pi} = 1.7;   &e^{eff}_{\nu} = 0.7 \\
pseudo SU(3): &e^{eff}_{\pi} = 2.2;  &e^{eff}_{\nu}= 1.2
\end{array}
\end{equation}
In summary, there are at least two different parametrizations of the
effective charges which can reproduce the experimental data. This result
weakens in some way the BNR conclusions. But it is indisputable that the
quadrupole moments in the normal parity sector are the same in the pseudo
SU(3) and the SSANM description. The seniority zero assumption in the
abnormal parity levels reduces the  total quadrupole moments predicted by
the pseudo SU(3) as
compared with the SSANM ones, resulting in greater
effective charges in any parametrization. We interpret this results
saying that  abnormal parity valence nucleons {\em adiabatically} follows
the behavior of the normal parity ones.

\section{Higher seniority states}

The single-j case of the SSANM was discussed by Eisenberg and Greiner
\cite{Eis70} as the extreme single-particle model. Given the
quadrupole operator $q_0$ does not mix states with different parity, the
abnormal parity state remain decoupled, and its matrix elements are
\cite{Rin80,deS63}
\begin{equation}
<l\frac 1 2 , jm|~q_0~|l\frac 1 2 ,jm>
= \{ j (j+1) - 3m^2 \} / {2j}                    \label{sj}
\end{equation}

\noindent
where  HO wave functions were used. This numbers are exactly
those given in Table IV of BNR for the abnormal parity states.

As a way to investigate the
spherical quasiparticle content of the abnormal sector,
let us analyze the contribution of two quasiparticle states,
coupled to angular momentum 2, to the quadrupole moment.
We use the Bogoliubov-Valatin transformation\cite{Rin80}
\begin{eqnarray}
 a^\dagger_{jm} = u_j \alpha^\dagger_{jm} + v_j \tilde\alpha_{jm} \\
\tilde a_{jm} = -v_j \alpha^\dagger_{jm} + u_j \tilde\alpha_{jm} \nonumber
\end{eqnarray}
with
\begin{equation}
\tilde a_{jm} = (-1)^{j+m} a_{j~-m}
\end{equation}
Given we are restricted to a state with $n_A$ particles in a single shell
with angular momentum
$j$, the $u$ and $v$ coefficients are

\begin{equation}
v =  \left\{ \frac {n_A} {2j+1} \right\} ^{1/2}, \hspace{.5cm}
u =  \left\{ \frac {2j+1-n_A} {2j+1} \right\} ^{1/2}
\end{equation}

\noindent
The two quasiparticle state with angular momentum 2 is
\begin{equation}
|j^n_A, 2M> = 2^{-1/2} [\alpha^\dagger_j \alpha^\dagger_j]^{2M}|0>
\end{equation}
\noindent
where $[...]^{JM}$ denotes angular momentum coupling and
$\alpha_{jm} |0> = 0 $.
 The quadrupole operator is expressed in second quantization as
\begin{eqnarray}
q_M = \sum_{j_1m_1 j_2m_2} <j_1m_1|q_M| j_2m_2> a^\dagger_{j_1m_1}
a_{j_2m_2}
\end{eqnarray}
\noindent
and, for a single $j$ and in terms of quasiparticle operators, becomes
\begin{equation}
q_M = \sqrt{ \frac {2j+1} 5 }<j\|q\|j> \{ uv[\tilde\alpha
\tilde\alpha]^{2M} -uv[\alpha^\dagger \alpha^\dagger ]^{2M} +
(u^2 - v^2)[\alpha^\dagger \tilde\alpha]^{2M} \}
\end{equation}

\noindent
The quadrupole moment is
\begin{eqnarray}
q_A(j) &= - \sqrt{2} \sqrt{\frac {2j+1} 5}<j\|q\|j> u v \\
       &= \{\frac {l(l+2)(2l+3)n_A(2l+2-n_A)}{5(l+1)(2l+1)} \}^{1/2}
\end{eqnarray}
for $n_A$ particles in a level with $j = l+1/2$

In Fig. 4 we have made a plot of $q_A(j)$ vs $n_A$ for the $h_{11/2}$ and
$i_{13/2}$ levels. In the same figure we plotted the quadrupole moments
obtained for the abnormal states in the SSANM, filling the single-j level with
pairs having the maximal possible quadrupole moments. It is apparent that
for two particles both results are quite similar (with two particles
you can have only states with seniority zero or two) but for any
other number
of particles the SSANM quadrupole moment is twice the two-quasiparticle
one, exhibiting a higher seniority content.
Giving such
dominance of higher seniority states in the ground state of even-even nuclei
imply the use of the seniority zero description of the
backbending is an oversimplification.

It is also interesting to analyze the quadrupole moment assigned in different
models to one particle in
the lowest energy state in the abnormal sector. This is the state with
lowest projection of the total angular momentum $m = \Omega = 1/2$.
In the single shell approach, this state has $j = \eta + 1/2$, and using
(\ref{sj}) its quadrupole moment is
$
q_0^{single j} = \eta (\eta + 2) / (2 \eta + 1) .
$
On the other side, in the asymptotic limit, i.e. allowing maximum mixing
between all the members of the $\eta$ shell with the same orbital angular
momentum $l$, the state $(\eta \eta 0) 1/2$ has, using (2), a quadrupole moment
$q_0^{asym} = 2 \eta$.
This asymptotic value is 3-4 times greater than the single-j one. The
fact that in both deformed Nilsson and Woods-Saxon potential the intruder
states remains essentially unmixed for deformations $\epsilon \approx
0.2-0.3$ \cite{Rin80,Ber89}
supports the single-j assignment. It also implies that the $\Omega = 1/2$ state
has not a constant slope, i.e. it has not reached its maximal quadrupole
moment at this deformation, as can be confirmed in Fig. 1.

In conclusion, we can say that
filling each single particle state with nucleons with their maximum
qua\-dru\-po\-le
moment is meaningful for those occupying normal parity states. However,
doing the same in the case of unique parity states contradicts microscopic
Nilsson and Woods-Saxon calculations.
It is reasonable to assume that nucleons in the unique parity
sector share the deformation of the others, and then they cannot be
described as pure seniority zero states.
The effective charges
used in the pseudo SU(3) description of nuclei includes the renormalization
of $\approx 25\%$
required for the fact the abnormal states remained ``frozen'' in the
model.

\section{Conclusions}

Some general features of the pseudo SU(3) model were presented.
Its predictive power for many  properties of heavy deformed nuclei was
exhibited and associated with its successful description of valence nucleons.
Core polarisation effects are included via the effective charges.

In the particular case of $BE2\uparrow$ transitions, the main ideas
proposed in BNR were analyzed. Their assumptions about the quadrupole
moments in the pseudo SU(3) formalism were rigorously checked, and found
correct as a first approximation. In the normal parity sector
 SU(3) states with largest deformation
are lowest in energy, and therefore result in the same states of maximal
deformation as in the SSANM.

In the abnormal parity sector the two and four quasiparticle content was
shown as a necessary interpretation of the contribution to quadrupole
moments of nucleons in these orbitals.
These quadrupole moments are non-zero, but only a fraction of their
asymptotic limit. The seniority zero approach is
exhibited as a very useful mathematical approach, which in general is
assigning nucleons in unique parity orbitals a passive role, but
needs greater effective charges to compensate for this assumption.

\section{Acknowledgments}
The authors want to thank the kind hospitality
received from Prof. Dr. W. Scheid
during their visit to the Institut fur Theoretische Physik,
JLU-Giessen, where much of this work was done.
Interesting comments from Dr. J. P. Draayer, Dr. S. Raman and
Dr. K. H. Bhatt are also acknowledged.

\end{section}

\newpage

\centerline{Table Captions}
\vskip 2cm

Table I: Proton and neutron occupation numbers in normal spaces
$n^N_\pi, n^N_\nu$, dominant irreps $(\lambda_\pi , \mu_\pi),
(\lambda_\nu , \mu_\nu)$, quadrupole moments $Q_\pi ,Q_\nu$ and
intrinsic
quadrupole moments $Q^{int}_\pi , Q^{int}_\nu$ for some rare earth and
actinide nuclei.

\vskip 1cm
Table II: The first column exhibits the
Nilsson labels $(\tilde N,\tilde n_z,\tilde \Lambda )\Omega^\pi $.
Quadrupole moments in the normal space evaluated in the pseudo
SU(3) and the SSANM are shown in the last two columns.

 \newpage
\centerline{Table I}
\vskip .5cm
\begin{eqnarray*}
\begin{array}{lrrrrrrrr}
Nuclei &n^N_\pi &n^N_\nu &(\lambda_\pi , \mu_\pi)
&(\lambda_\nu , \mu_\nu) &Q_\pi &Q^{int}_\pi &Q_\nu &Q^{int}_\nu\\
\\
^{154}Sm &6 &6 &(12,0) &(18,0) &31.5 &30.0 &45.3 &43.2\\
^{156}Gd &8 &6 &(10,4) &(18,0) &32.3 &30.0 &44.5 &43.2\\
^{160}Dy &10 &8 &(10,4) &(18,4) &31.9 &30.0 &44.5 &43.2\\
^{164}Dy &10 &10 &(10,4) &(20,4) &31.9 &30.0 &54.5 &52.8\\
^{166}Yb &12 &8 &(12,0) &(18,4) &30.9 &30.0 &50.7 &48.0\\
^{168}Yb &12 &10 &(12,0) &(20,4) &30.9 &30.0 &55.5 &52.8\\
^{174}Yb &12 &14 &(12,0) &(20,6) &30.4 &30.0 &58.3 &55.2\\
^{232}Th &4 &10 &(12,2) &(30,4) &32.3 &31.2 &77.1 &74.7\\
^{234}U &6 &10 &(18,0) &(30,4) &44.2 &43.2 &77.2 &74.7\\
^{238}U &6 &12 &(18,0) &(36,0) &44.4 &43.2 &86.3 &84.8\\
^{240}U &6 &14 &(18,0) &(34,6) &43.9 &43.2 &89.1 &86.3
\end{array}
\end{eqnarray*}

\vskip 1cm
\centerline{Table II}
\vskip .5cm
\begin{eqnarray*}
\begin{array}{ccr}
(\tilde N,\tilde n_z,\tilde \Lambda )\Omega^\pi
 &q^{\tilde{SU(3)}}_\Omega  &q^{SSANM}_\Omega   \\
\\
(3,3,0){\frac 1 2}^+ &7.5 &7.4\\
(3,2,1){\frac 1 2}^+ &3.7 &3.8\\
(3,2,1){\frac 3 2}^+ &3.7 &3.5\\
(3,1,2){\frac 3 2}^+ &0.0  &0.2\\
(3,1,2){\frac 5 2}^+ &0.0  &-0.2\\
(3,1,0){\frac 1 2}^+ &0.0  &-0.2\\
\\
(4,4,0){\frac 1 2}^- &9.6 &9.6\\
(4,3,1){\frac 1 2}^- &6.0 &6.2\\
(4,3,1){\frac 3 2}^- &6.0 &5.9\\
(4,2,2){\frac 3 2}^- &2.4 &2.7\\
(4,2,2){\frac 5 2}^- &2.4 &2.2\\
(4,2,0){\frac 1 2}^- &2.4 &2.2
\end{array}
\end{eqnarray*}

\newpage
\centerline{Figure Captions}

\vskip 2cm
Fig. 1. Nilsson level scheme for the $\eta =5 (\tilde\eta =4)$ proton
shell. At deformation $\epsilon = 0$ the levels are labeled with
spherical shell model quantum numbers, while at $\epsilon \ne 0$ the
asymptotic pseudo quantum numbers labels $\tilde\Omega [\tilde\eta \tilde
n_z \tilde\Lambda ]$ are given.

\vskip 1cm
Fig. 2. Energy spectra for $^{126}Ba$. On the right hand side are spectra
obtained diagonalizing separately the interaction in the $\tilde f \tilde
p \  (\tilde\lambda ,\tilde\mu )=(24,0)$ and the $h_{11/2}, v=0,2$ spaces.
In the center is the composite result which includes the effect of the
interaction between the normal-parity and the $h_{11/2}$ nucleons.

\vskip 1cm

Fig. 3. Quotient between theoretical and experimental Q values for
$^{168}Er$, for the pseudo SU(3)(solid line) and the SSANM (dashed
line). In Fig. 1a the parametrization (\ref{Ram1}) was used, while Fig.
1b refers to (\ref{nos}).

\vskip 1cm
Fig. 4. The total quadrupole deformation in the abnormal parity space
plotted against the number $n_A$ of particles in the abnormal space. The
solid line refers to the two quasiparticle estimation, while the dashed
lined refers to the SSANM description. Fig. 4a(4b) was performed to a
single shell with j=11/2 (j=13/2).


\begin{thebibliography}{Feni93}
\bibitem{Rat73} R.D. Ratna Raju, J.P. Draayer and K. T. Hecht, Nucl.
Phys.{\bf A202} (1973) 433.
\bibitem{Rat79} R.D. Ratna Raju, K. T. Hecht, B. D. Chang and J. P.
Draayer, Phys. Rev. {\bf C20} (1979) 2397.
\bibitem{Dra81} J. P. Draayer, C.S. Han, K. T. Weeks and K. T. Hecht,
Nucl.
Phys. {\bf A365} (1981) 127; K. J. Weeks, J. P. Draayer and C. S. Han,
Nucl. Phys. {\bf A371} (1981) 19.
\bibitem{Dra84}J. P. Draayer and K. J. Weeks, Ann. Phys. {\bf 156}
(1984) 41
\bibitem{Ari69} A. Arima, M. Harvey and K. Shimizu, Phys. Lett {\bf B30}
(1969) 517; K. T. Hecht and A. Adler, Nucl. Phys. {\bf A137} (1969) 129.
\bibitem{Cas87} O. Casta\~nos, J.P. Draayer and Y. Leschber, Ann. Phys
{\bf 180} (1987) 290.
\bibitem{Hir92} J. G. Hirsch, O. Casta\~nos, P. O. Hess, Proc. Franklin
Symposium in Celebration of the Discovery of the Neutrino, April 1992 (World
Scientific, Singapore); O. Casta\~nos, J. G. Hirsch, P. O. Hess, Rev. Mex.
Fis. {\bf 39} Supl. 2 (1993) 29; O. Casta\~nos, J. G. Hirsch, O. Civitarese,
P. O. Hess, Nucl. Phys. {\bf A} in press.
\bibitem{Rin80}P. Ring and P. Schuck, ``The Nuclear Many-Body Problem",
Springer-Verlag N.Y. 1980.
\bibitem{Ber89} R. Bergtsson, J. Dudek, W. Nazarewicz and P. Olanders,
Phys. Scripta {\bf 39} (1989) 196.
\bibitem{Bha92} K.H. Bhatt, C.W. Nestor,Jr. and S. Raman; Phys. Rev
{\bf C46}(1992)164.
\bibitem{Ell58} J. P. Elliott, Proc. Roy. Soc. {\bf A245} (1958) 128,
562; J. P. Elliott and M. Harvey, Proc. Roy. Soc.{\bf A272} (1963) 557.
\bibitem{Str72}D. R. Strottman, Nucl. Phys. {\bf A 188} (1972) 488.
\bibitem{Hec75} K. T. Hecht, Phys. Lett. {\bf B58} (1975) 253.
\bibitem{Bra78} D. Braunschwieg and K. T. Hecht, Phys. Lett. {\bf B77}
(1978) 33.
\bibitem{Cas92} O. Casta\~nos, M. Moshinsky and C. Quesne, Phys. Lett.
{\bf B277} (1992) 238.
\bibitem{Dra82} J. P. Draayer, K. J. Weeks and K. T. Hecht, Nucl. Phys.
{\bf A381} (1982) 1.
\bibitem{Naq90} H. A. Naqvi and J.P. Draayer, Nucl. Phys. {\bf A516}
(1990) 351.
\bibitem{Ste72} F. S. Stephens and R. S. Simon, Nucl. Phys. {\bf A183}
(1972) 257.
\bibitem{Dra73} Y. Akiyama and J. P. Draayer, Comp. Phys. Commun. {\bf 5}
(1973) 405, J. P. Draayer, Y. Leschber, S. C. Park and R. Lopez,
Comp. Phys. Commun. {\bf 56} (1989) 279.
\bibitem{Ram91} S. Raman, C.W. Nestor,Jr., S. Kahane and K. H. Bhatt,
Phys. Rev. {\bf C43} (1991) 556.
\bibitem{Boh69}A. Bohr and B.R. Mottelson,``Nuclear Structure" V.I, W.A.
Benjamin Inc. N.Y., 1969.
\bibitem{Rin73}P. Ring, R. Bauer and J. Speth, Nucl. Phys. {\bf A206}
(1973) 97.
\bibitem{Kis63} L.S. Kisslinger and R. A. Sorensen, Rev. Mod. Phys. {\bf
35} (1963) 853.
\bibitem{Eis70} J. M. Eisenberg and W. Greiner, ``Nuclear Models" V. 1,
North-Holland Publ. Co. Amsterdam $3^{rd}$ Ed. 1987.
\bibitem{deS63} A. de-Shalit and I. Talmi, ``Nuclear Shell Theory",
Academic Press N.Y. 1963.
\end{thebibliography}
\end{document}